\documentclass[twocolumn,showpacs,nofootinbib,amsmath,amssymb,aps,prd,superscriptaddress,balancelastpage]{revtex4}

\usepackage[dvips]{graphicx}
\usepackage{graphicx}
\usepackage{epsf}
\usepackage{amsmath}
\usepackage{amssymb}

\usepackage{graphicx}
\usepackage{dcolumn}
\usepackage{bm}
\usepackage{graphicx}
\usepackage[caption=false]{subfig}
\usepackage{amsmath,amssymb}
\usepackage{amsfonts,amssymb,mathrsfs}
\usepackage[bookmarks=false,pdfstartview=FitH]{hyperref}

\def\be{\begin{equation}}
\def\ee{\end{equation}}
\def\bea{\begin{eqnarray}}
\def\eea{\end{eqnarray}}

\usepackage{color}

\begin{document}

\title{Emergent Universe and Genesis from the DHOST Cosmology}

\author{Amara Ilyas}
\affiliation{Department of Astronomy, School of Physical Sciences, University of Science and Technology of China, Hefei, Anhui 230026, China}
\affiliation{CAS Key Laboratory for Researches in Galaxies and Cosmology, University of Science and Technology of China, Hefei, Anhui 230026, China}
\affiliation{School of Astronomy and Space Science, University of Science and Technology of China, Hefei, Anhui 230026, China}

\author{Mian Zhu}
\affiliation{Department of Physics, The Hong Kong University of Science and Technology, Clear Water Bay, Hong Kong S.A.R., China}
\affiliation{HKUST Jockey Club Institute for Advanced Study, The Hong Kong University of Science and Technology, Clear Water Bay, Hong Kong S.A.R., China}

\author{Yunlong Zheng}
\email{zhyunl@ustc.edu.cn}
\affiliation{Department of Astronomy, School of Physical Sciences, University of Science and Technology of China, Hefei, Anhui 230026, China}
\affiliation{CAS Key Laboratory for Researches in Galaxies and Cosmology, University of Science and Technology of China, Hefei, Anhui 230026, China}
\affiliation{School of Astronomy and Space Science, University of Science and Technology of China, Hefei, Anhui 230026, China}

\author{Yi-Fu Cai}
\email{yifucai@ustc.edu.cn}
\affiliation{Department of Astronomy, School of Physical Sciences, University of Science and Technology of China, Hefei, Anhui 230026, China}
\affiliation{CAS Key Laboratory for Researches in Galaxies and Cosmology, University of Science and Technology of China, Hefei, Anhui 230026, China}
\affiliation{School of Astronomy and Space Science, University of Science and Technology of China, Hefei, Anhui 230026, China}


\begin{abstract}
In this article, we present an emergent universe scenario that can be derived from DHOST cosmology. The universe starts asymptotically Minkowski in the far past just like the regular Galileon Genesis, but evolves to a radiation dominated period at the late stage, and therefore, the universe has a graceful exit which is absent in the regular Galileon Genesis.
We analyze the behavior of cosmological perturbations and show that both the scalar and tensor modes are free from the gradient instability problem.
We further analyze the primordial scalar spectrum generated in various situations and discuss whether a scale invariance can be achieved.

\end{abstract}

\maketitle

\newcommand{\eq}[2]{\begin{equation}\label{#1}{#2}\end{equation}}

\section{Introduction}

Alternative scenarios to inflationary cosmology are of
interest since it can explain the formation of the Large Scale Structure of our universe as good as inflation (for example, see \cite{Brandenberger:2009jq, Brandenberger:2011gk} for comprehensive reviews on theoretical paradigms of the very early universe alternative to inflation). The alternative scenarios can also avoid the initial spacetime singularity, a generic problem in the inflationary cosmology \cite{Borde:1993xh,Borde:2001nh}. The removal of initial singularity is widely acknowledged in the framework of bounce cosmologies (see \cite{Mukhanov:1991zn, Brandenberger:1993ef, Cai:2008qw, Cai:2012va, Yoshida:2017swb} and some recent reviews in \cite{Novello:2008ra, Lehners:2008vx, Cai:2014bea, Battefeld:2014uga, Brandenberger:2016vhg, Cai:2016hea}). There is another paradigm of the very early universe, dubbed as the emergent universe scenario \cite{Ellis:2002we, Ellis:2003qz}, in which the universe is emergent from a quasi-Minkowski spacetime. The scenario was first postulated in the string gas cosmology \cite{Brandenberger:1988aj}, in which a specific pattern for primordial cosmological perturbations has been predicted \cite{Nayeri:2005ck, Brandenberger:2006xi, Brandenberger:2006vv, He:2016uiy} (also see \cite{Battefeld:2005av, Brandenberger:2011et, Brandenberger:2015kga} for recent reviews).

Recently, the proposal of the conformal Galilean model \cite{Nicolis:2008in} has inspired some particular alternative to inflation cosmologies, such as G-bounce \cite{Qiu:2011cy, Easson:2011zy} and Galilean Genesis \cite{Creminelli:2010ba}. In particular, the model of Galilean Genesis describes that the universe starts from the Minkowski spacetime, which corresponds to a specific configuration of the Galilean field with zero energy density. This configuration is not stable \cite{Libanov:2016kfc}, and even a small classical perturbation can drive the universe to deviate from the original state. Therefore, one requires that the Galilean field has to decay into radiation very quickly before the universe reaches the big rip singularity \cite{LevasseurPerreault:2011mw}. Moreover, the perturbations of the Galilean field could propagate superluminally in the original model. Hence, several generalized Galilean Genesis models have been developed in the literature \cite{Creminelli:2012my, Hinterbichler:2012yn, Nishi:2014bsa, Nishi:2015pta, Mironov:2019qjt}. However, in the model of subluminal Galileon Genesis, there will always exist a region of phase space where the perturbations propagate superluminally (with arbitrarily high speed) when one includes any matter which is not directly coupled to the Galileon \cite{Easson:2013bda}. Furthermore, the gradient instabilities appear during the transition from the Genesis phase to the epoch immediately after \cite{Kobayashi:2015gga}.

In this paper, we explore the realization of the emergent universe scenario depicted by the Degenerate Higher-Order Scalar-Tensor (DHOST) theory \cite{Langlois:2015skt, BenAchour:2016fzp,Langlois:2017mxy}. Our model can gracefully exit the emergent phase and transfer to a radiation dominated period, without triggering various instabilities. In this model, we deform the kinetic term for the background scalar field so one could approximately approach the Galilean symmetry in the far past. The symmetry then is manifestly broken along with the evolution of the scalar field. Consequently, the big rip singularity disappears, and the state of Genesis is replaced by a smooth process of the fast roll expansion. Eventually, the universe can evolve into the radiation-dominated phase with appropriate parameter choices. 

However, the sound speed squared $c_s^2$, which controls the propagation of primordial perturbations, is found to become negative for a short while when the universe exits the state of quasi-Minkowski spacetime, as encountered by the original Galileon Genesis model as well as generalized cases \cite{Libanov:2016kfc}. To address this issue, we extend the Galileon action to the form suggested by the DHOST theory, a natural extension to the Galileon theory. We find that $c_s^2$ can be positive in the whole cosmological evolution with the DHOST coupling.

The present paper is organized as follows. We introduce the cosmology of the previously studied Galilean Genesis models in Section \ref{sec:GGcosmology}, and propose the improved version of Galilean Genesis (DHOST genesis) in Section \ref{sec:DHOSTgenesis} . We study the background dynamics of the universe for our model in Section \ref{sec:background}. We provide a detailed perturbation analysis in Section \ref{sec:Fluctuations}, where the stability of the scalar perturbation is demonstrated, and also discuss the primordial spectrum of scalar perturbations generated by vacuum fluctuations and thermal fluctuations. We conclude with a discussion in Section \ref{sec:conclude}.

Throughout this paper, we take the sign of the metric to be (+,-,-,-), the canonical kinetic term is defined as $X \equiv \frac{1}{2}\nabla_{\mu} \pi \nabla^{\mu} \pi$, and the reduced Planck mass is taken to be one, i.e., $M_p^2 \equiv 1/(8\pi G) = 1$.


\section{Cosmology of (Sub-luminal) Galilean Genesis} \label{sec:GGcosmology}

We briefly review the cosmology of Galilean Genesis, which is realized by a scalar field minimally coupled to Einstein gravity. The Lagrangian is given by \cite{Creminelli:2010ba}
\begin{eqnarray}\label{L1}
 {\cal L} = -f^2 e^{2\pi} (\partial \pi)^2 + \frac{\beta}{2} \gamma (\partial \pi)^4 + \gamma (\partial \pi)^2  \Box \pi ~,
\end{eqnarray}
where the scalar field $\pi$ is dimensionless. We define the operator $\Box \equiv \nabla_\mu\nabla^\mu$. For the original model of Galilean Genesis, the coefficients are set as $\beta = 1$ and $\gamma = \frac{f^3}{\Lambda^3}$, where $f$ and $\Lambda$ are two model parameters of mass dimension. The Lagrangian has a negative sign in front of the regular kinetic energy, which is represented by the first term of Eq.\eqref{L1}. The second term, which is a positive definite higher-order kinetic energy term, stabilizes the model. Thus, the combination of the first and second term in Eq. \eqref{L1} exhibits the property of the ghost condensate \cite{ArkaniHamed:2003uy}. This type of Lagrangian enjoys the internal Galilean invariance $\pi\rightarrow \pi+b_\mu x^\mu$, which has been observed in \cite{Creminelli:2012my}. 

When applied into the cosmological background, the above Lagrangian yields the energy density and the pressure
\begin{align*}
 \rho_{\pi} &= -e^{2\pi} f^2 \dot\pi^2 +\frac{3}{2}\beta\gamma \dot\pi^4 +6\gamma H\dot\pi^3 ~, \\
 p_{\pi} &= -e^{2\pi} f^2 \dot\pi^2 +\frac{1}{2}\beta\gamma \dot\pi^4 -2\gamma\dot\pi^2\ddot\pi ~,
\end{align*}
and the equation of motion(EoM) can be obtained by the conservation equation. By solving the EoM, one can find an interesting solution of a static Minkowski universe if there is,
\begin{align}\label{eq:pi}
 \pi = \ln(-\frac{1}{M_G t}) ~,~~ M_G^2 \equiv \frac{2f^2}{3\beta \gamma} ~.
\end{align}
In this solution, it is easy to check that the energy density of the Galilean field $\rho_\pi$ vanishes, and there is no expansion of the universe. However, the pressure evolves as 
\begin{align*}
 p_{\pi} = - \frac{(2+\beta)\gamma}{t^4} ~,
\end{align*}
which is negative and non-vanishing. Therefore, the above solution corresponds to the emergent universe scenario. 

In the above model, the fluctuations propagate superluminally around the NEC-violating background. Thus, an updated version of the subluminal Galilean Genesis was put forward in \cite{Creminelli:2012my}, where the coefficient $\beta$ is modified from unity to be an arbitrary constant, i.e., $\beta$ is required to be between $1$ and $4$ so that the model is free of instability and superluminality. However, when considering other matter components in a more realistic universe, that are not directly coupled to the background field, it was shown in \cite{Easson:2013bda} that there always exists a phase space suffering from the superluminality problem. The result is tightly related to the fact that the Galilean field cannot exit the pre-emergent phase gracefully unless it has to be assumed to decay into other fields through a defrosting phase \cite{LevasseurPerreault:2011mw}.

\section{Improved DHOST Genesis} \label{sec:DHOSTgenesis}

To address the general issue existing in cosmologies of Galilean Genesis, we would like to improve the model by designing a graceful exit mechanism for the Galilean field. An updated version of Galilean Genesis Lagrangian without external matter is given by
\begin{eqnarray} \label{eq:GGv1}
{\cal L}_\pi = -f^2 g(\pi) (\partial \pi)^2 + \frac{\beta}{2} \gamma (\partial \pi)^4 + \gamma (\partial \pi)^2  \Box \pi ~,
\end{eqnarray}
where the coefficients $\beta$ and $\gamma$ are the model parameters, same as the previous Galilean Genesis model. Compared with Eq. \eqref{L1}, the only difference is the appearance of a function $g(\pi)$ suggested to be
\begin{align*}
 g(\pi) = \frac{e^{-2\pi}-e^{2\pi}}{e^{-4\pi}+e^{4\pi}} ~.
\end{align*}

\begin{figure}
\includegraphics[scale=0.34]{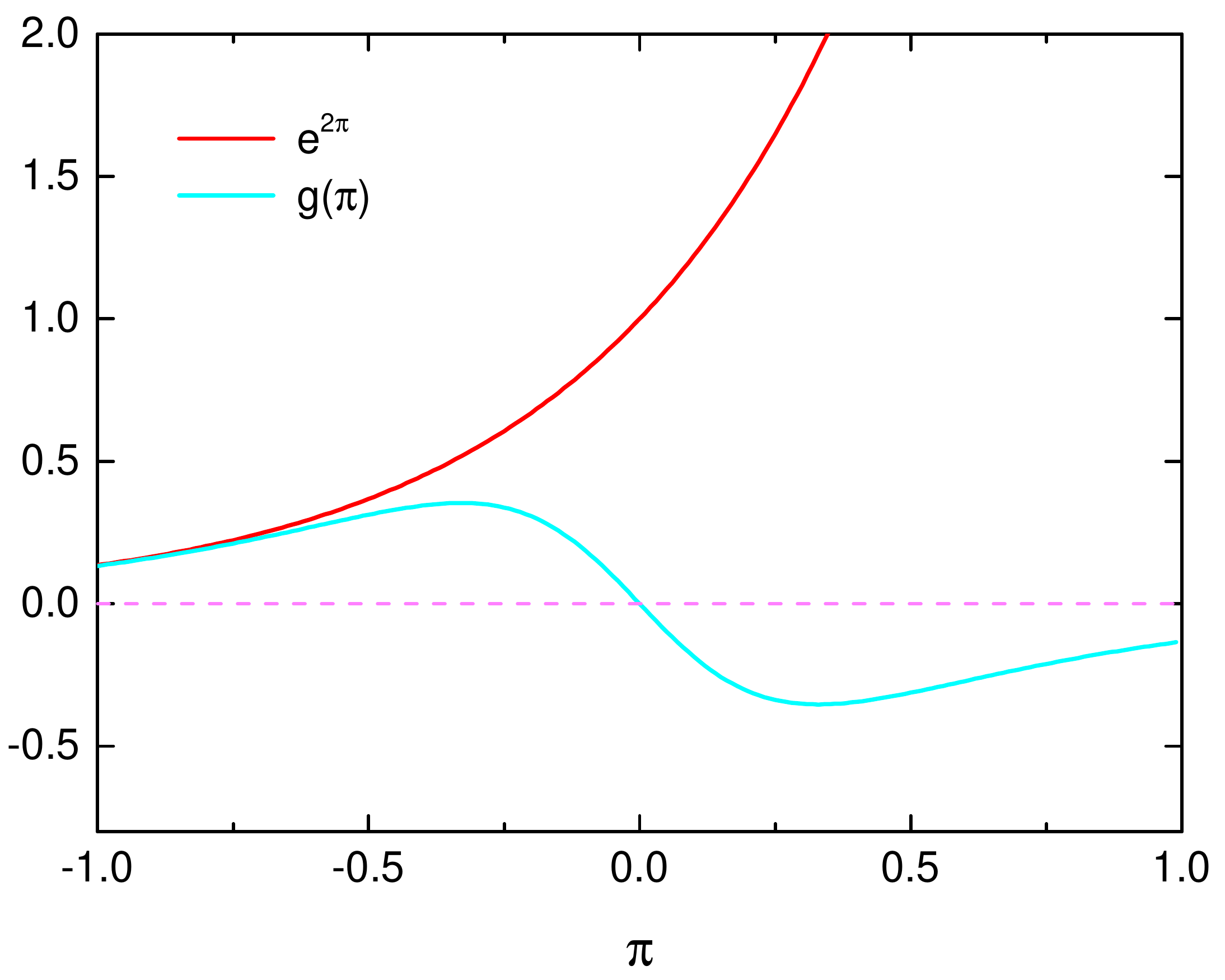}
\caption{Plots of the coefficient $g$ as a function of the scalar field $\pi$. The solid blue curve depicts the form of $g(\pi)$ in the Eq. \eqref{eq:GGv1}, and the solid red line represents the traditional model of Galilean Genesis as described by Eq. \eqref{L1}.}
\label{Fig1}
\end{figure}

It is easy to see that $g(\pi)$ goes to $e^{2\pi}$ when $\pi \ll -1$, and the traditional Lagrangian of Galilean Genesis is recovered in this limit. In this regime, the sign in front of the conventional kinetic term $(\partial \pi)^2$ is negative, and correspondingly, the model shares the property of the ghost condensate due to the inclusion of the positive definite term $(\partial \pi)^4$. However, when $\pi \gg 1$, $g(\pi)$ approaches to $-e^{-2\pi}$ and therefore yields a sign change in front of the conventional kinetic term. If $\pi$ can evolve to the region $\pi > 0$, the universe can transfer from the emergent universe phase to an expanding phase. We depict the function $g(\pi)$ and the traditional choice $g(\pi) \equiv e^{2\pi}$ in Fig. \ref{Fig1} to demonstrate their difference.

However, the improved form of Lagrangian in Eq. \eqref{eq:GGv1} does not solve the gradient instability problem. Thus, we extend the scenario into the DHOST theory \cite{Langlois:2015skt,BenAchour:2016fzp, Langlois:2017mxy} in order to improve the behavior of $c_s^2$. We give a brief introduction to the DHOST theory in appendix \ref{app:DHOSTintro},

The action for our updated DHOST Genesis model is
\begin{eqnarray} \label{eq:bounceS}
 S &= \int d^4x \sqrt{-g} \Big[ - \frac{1 + h}{2} R + K(\pi,X) + G(\pi,X) \Box \pi \nonumber \\ 
 & - \frac{h}{4X} \big( L_1^{(2)} - L_2^{(2)} \big) + \frac{h-2Xh_X}{4X^2} \big( L_4^{(2)} - L_3^{(2)} \big) \Big] ~,
\end{eqnarray}
where the first term $-{R}/{2}$ corresponds to the standard Einstein-Hilbert action, and $K(\pi, X) + G(\pi, X) \Box \pi$ is the Galileon terms, whose detailed form is taken to be:
\begin{align*}
 K(\pi,X) =  -2 f^2 g(\pi) X+ 2 \beta \gamma X^2~,~~ G(\pi,X) = 2 \gamma X  ~. 
\end{align*}
Finally, $h \equiv h(X)$ is a function of $X$ only which represents the DHOST coupling, and $h_X \equiv (\partial h)/(\partial X) $. In our case $h(X)$ is simply taken to be
\begin{align*}
h(X) = d_1 X+d_2 X^2 ~,
\end{align*}
and the DHOST Lagrangian is defined as
\begin{align*} 
& L_1^{(2)} = \pi_{\mu \nu} \pi^{\mu \nu} ~,~ L_2^{(2)} = (\Box \pi)^2 ~,  \nonumber\\ 
& L_3^{(2)} =  (\Box \pi)\pi^{\mu}\pi_{\mu \nu} \pi^{\nu} ~,~ L_4^{(2)} = \pi_{\mu}\pi^{\mu \rho}\pi_{\rho \nu}\pi^{\nu} ~.
\end{align*}

The action \eqref{eq:bounceS} does not contain any Ostrogradski ghost degree of freedom, and in the next two sections, we shall analyze the cosmological evolution and show the absence of instabilities as well.

\section{Background Dynamics} \label{sec:background}

In this section, we work with the background dynamics of our improved DHOST Genesis Model. We present the dynamical equation and analysis of the stability issue at the background level in Subsection \ref{subsec:bgeom}. After that, we solve the background dynamics numerically and give a parameterization in Subsection \ref{subsec:bgnumerical}. 

\subsection{Equation of motions for background dynamics and its stability issue} \label{subsec:bgeom}

We consider a spatially flat FRW background 
$ds^2=dt^2-a^2(t)d\vec{x}^2$ 
and calculate the Friedman equations by the variation of Eq. \eqref{eq:bounceS} with respect to the metric as
\begin{align}
\label{eq:Friedmann1}
 3H^2 &= - f^2 g(\pi) \dot\pi^2 + \frac{3}{2}\beta\gamma\dot\pi^4 + 6 \gamma H \dot\pi^3 ~, \\
\label{eq:Friedmann2}
 -2\dot{H} - 3H^2 &= - f^2 g(\pi) \dot\pi^2 + \frac{1}{2}\beta\gamma\dot\pi^4 - 2 \gamma \dot\pi^2 \ddot\pi ~.
\end{align}
Note that the function $h(X)$ does not appear in the background Eq. \eqref{eq:Friedmann1} and Eq. \eqref{eq:Friedmann2}, so the DHOST term has no contribution to the background dynamics. Thus, the analysis of Galileon Genesis is valid in our model.

We may get the energy density $\rho_{\pi}$ and the pressure $p_{\pi}$ by the Friedmann equation $3H^2 = \rho_{\pi}$ and $-2\dot{H} = \rho_{\pi} + p_{\pi}$, which implies the equation of state(EoS) parameter to be $ w_\pi \equiv p_\pi / \rho_\pi ~$. We also define a useful parameter $\epsilon_H=-\frac{\dot{H}}{H^2}$, which characterizes the background evolution.

In addition, we can write down the generalized Klein-Gordon equation by varying the Lagrangian with respect to the scalar field $\pi$ as
\begin{eqnarray}\label{eom_pi}
 {\cal P} \ddot\pi + {\cal F}\dot\pi = 0~,
\end{eqnarray}
where the form of ${\cal F}$ is given by
\begin{align}\label{cal F}
{\cal F} = & - 6 \beta \gamma^2\dot\pi^5 - 18 \gamma^2 H \dot\pi^4 + 6 f^2 \gamma g(\pi)  \dot\pi^3  + 6 \beta \gamma H \dot\pi^2 \nonumber\\
& + 18 \gamma H^2 \dot\pi - f^2 g_{,\pi}\dot\pi - 6 f^2 g(\pi) H ~,
\end{align}
which corresponds to the friction term of the generalized Klein-Gordon equation. The subscript ``$_{,\pi}$" denotes the derivative with respect to $\pi$. For a canonical scalar field, the friction term is $3H$, which can be read from the last term of Eq. \eqref{cal F}, when $f^2g = -1/2$. However, in our case, this term also depends on other parameters, namely, the time derivative of the scalar field $\dot\pi$, the Hubble expansion rate $H$, and its time derivative $\dot{H}$. Note that, in the above expression, we have already used the second Friedmann equation. If there is any other matter component presented in the Friedmann equation, the expression for $\mathcal{F}$ will also depend on the energy density and pressure of the other matter field. 

The other parameter $\cal {P}$, which appears in front of the term $\ddot\pi$ in Eq. \eqref{eom_pi}, is expressed as
\begin{align*}\label{eq:P}
{\cal P} = 6 \gamma^2 \dot\pi^4 + 6 \beta \gamma \dot\pi^2 + 12 \gamma H \dot\pi - 2 f^2 g(\pi) ~, 
\end{align*}
and characterizes the positivity of the kinetic energy of the model. If $\mathcal{P}$ is negative, then there will be a ghost mode with energy state unbounded from below, leading to quantum instability. Thus, we expect $\mathcal{P}$ to be positive definite throughout the cosmological evolution. It is obvious that $\mathcal{P}$ is positive at the high energy regime due to the presence of $\dot{\pi}^4$ term with $\dot{\pi}$ to be large enough. A large $\dot{\pi}$ exists during the whole pre-emergent phase, where the higher derivative terms are dominant. Afterward, if the universe exits the pre-emergent phase with a sign change in front of the $\dot{\pi}^2$ term, then the function $\mathcal{P}$ can keep being positive at low energy scales. Therefore, the model can be free of the ghost issue, and this will be verified later in detailed numerical analysis.

\subsection{Numerical evaluation and parameterization} \label{subsec:bgnumerical}

\begin{figure*}[th]
	\centering
	\subfloat[]
	{\includegraphics[width=0.33\textwidth]{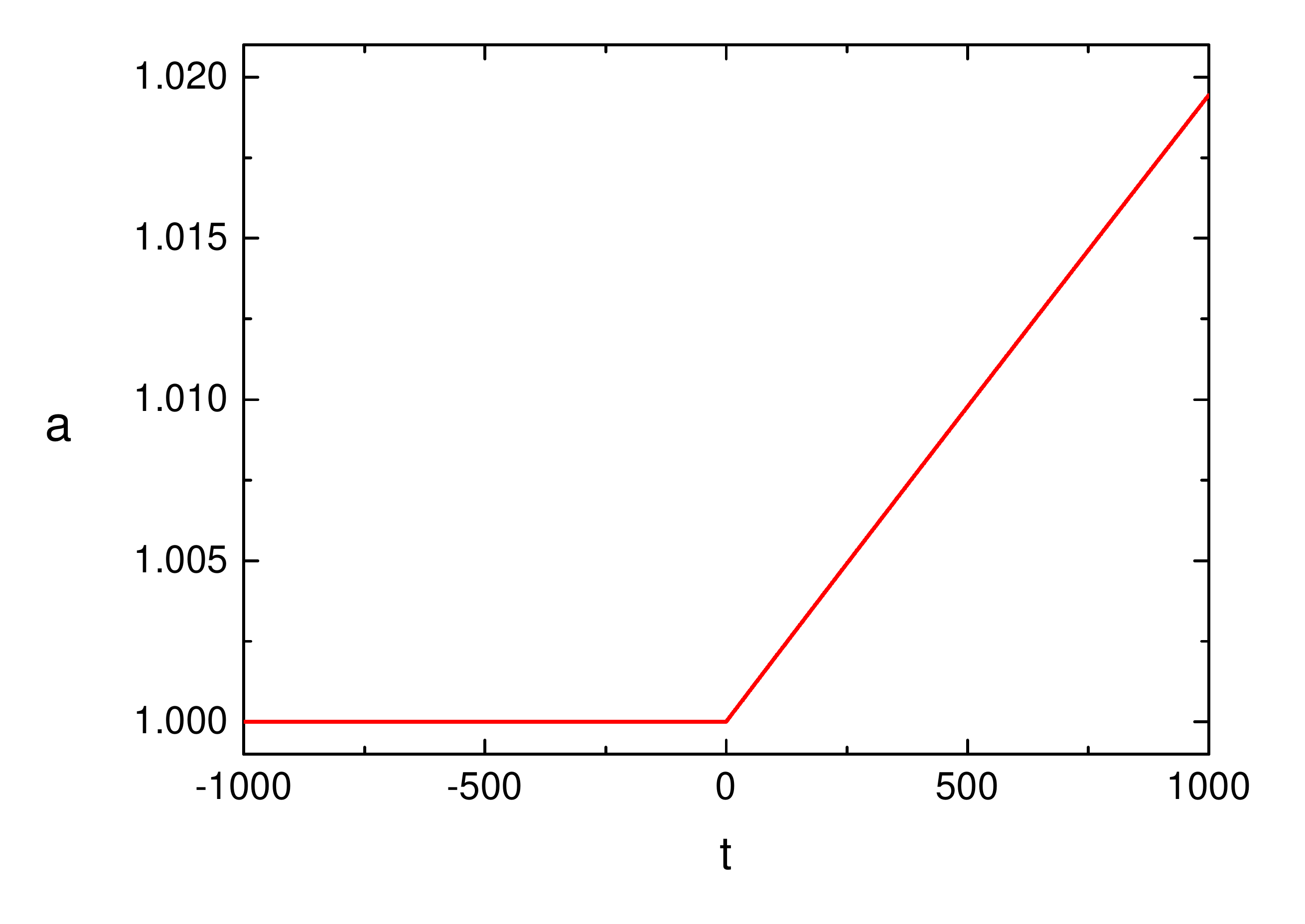} \label{fig:bga}}
	\subfloat[]
	{\includegraphics[width=0.34\textwidth]{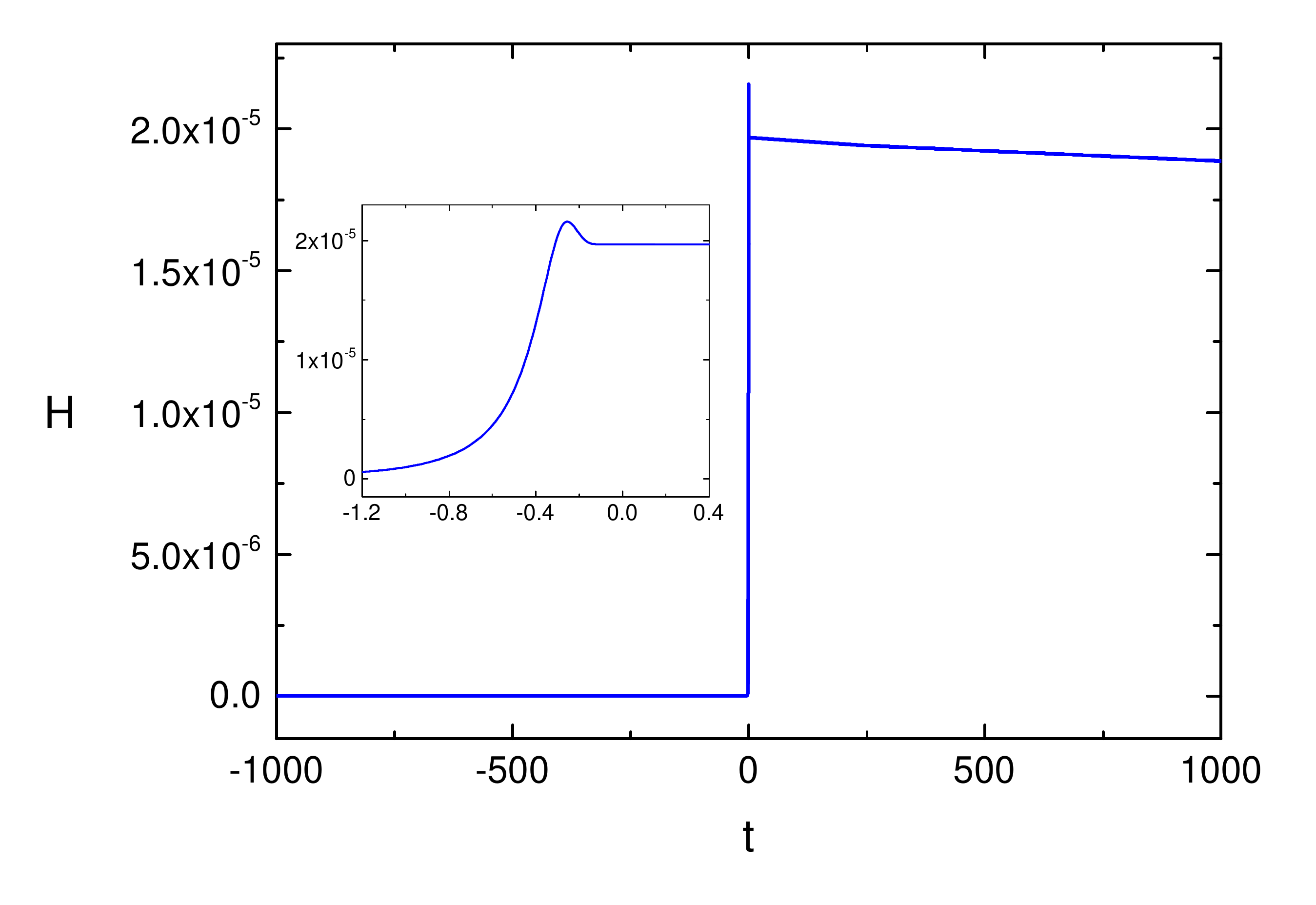} \label{fig:bgH}}
	\subfloat[]
	{\includegraphics[width=0.32\textwidth]{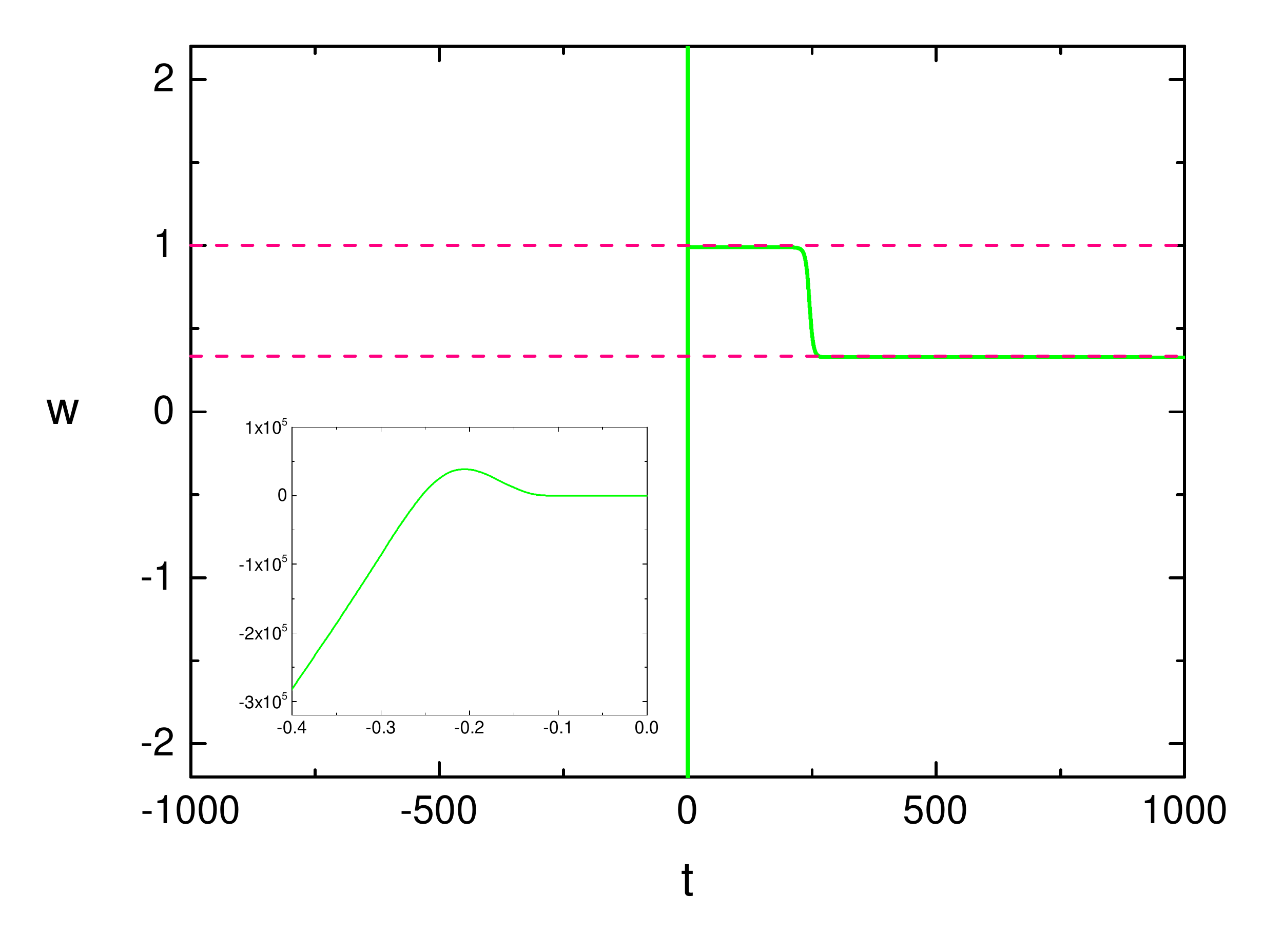} \label{fig:bgw}}
	\caption[]{\footnotesize \hangindent=10pt
		\it{Evolution of the background geometry as a function of cosmic time $t$. The scale factor $a$, the Hubble parameter $H$, and the EoS parameter $w$ is plotted by red, blue, and green solid line, respectively. The approximate behavior of $w$, i.e. $w \simeq 1$ shortly after the genesis and $w \simeq 1/3$ for the far future, is exhibited by the pink dash line.}
	}
	\label{fig:bggeometry}
\end{figure*}

\begin{figure*}[th]
\centering
\subfloat[]
{\includegraphics[width=0.32\textwidth]{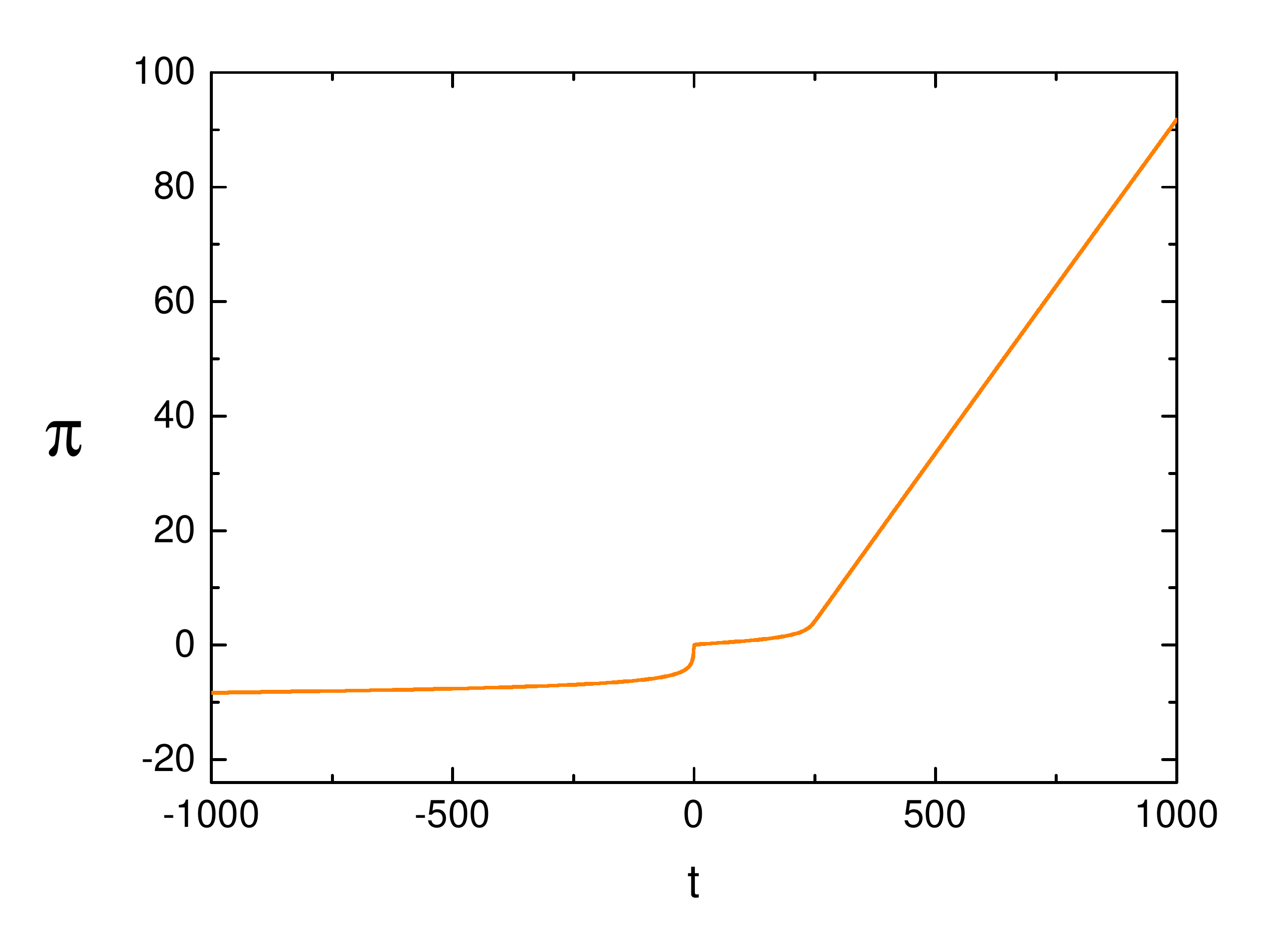} \label{fig:bgpi}}
\subfloat[]
{\includegraphics[width=0.32\textwidth]{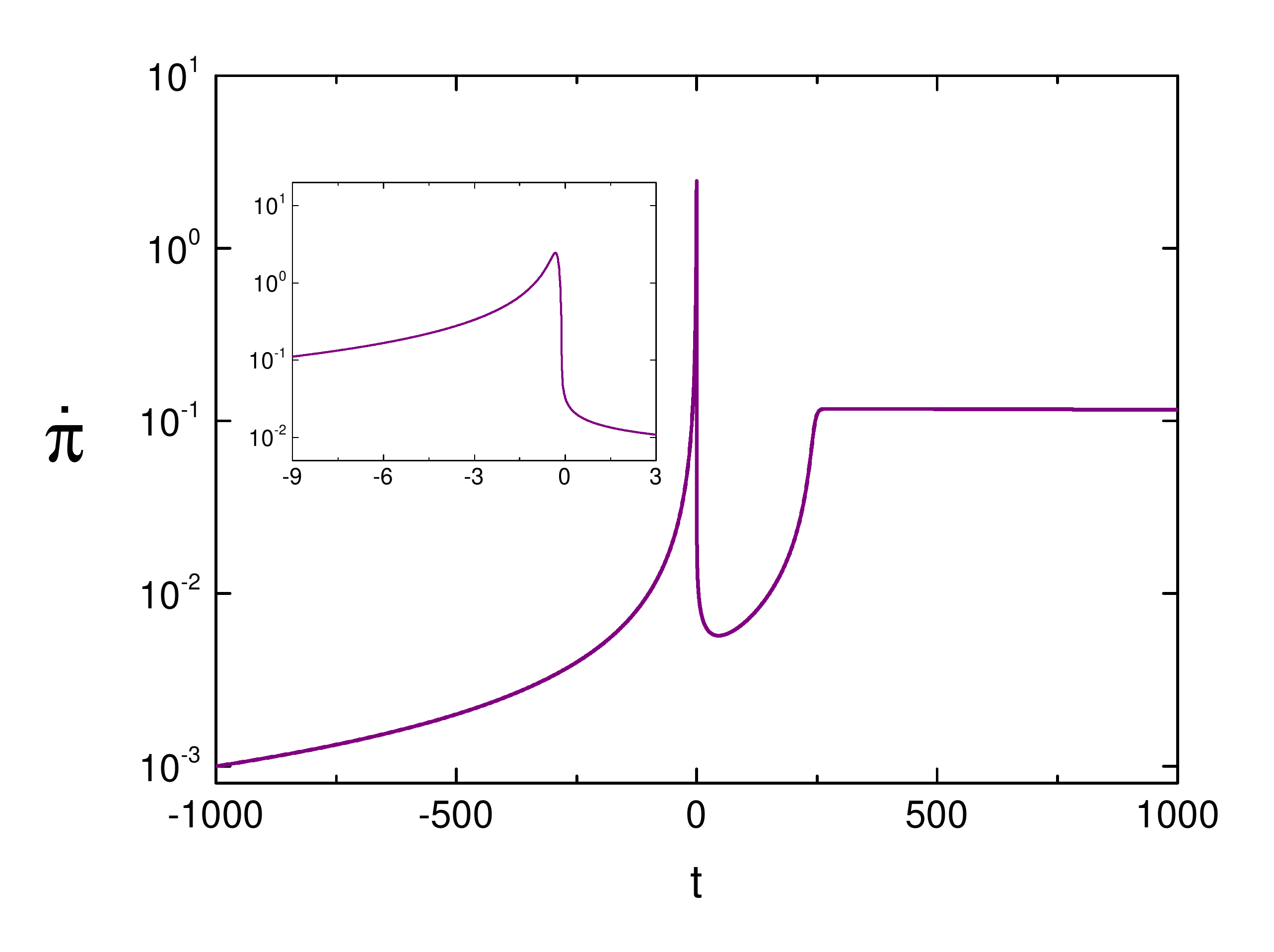} \label{fig:bgpit}}
\subfloat[]
{\includegraphics[width=0.31\textwidth]{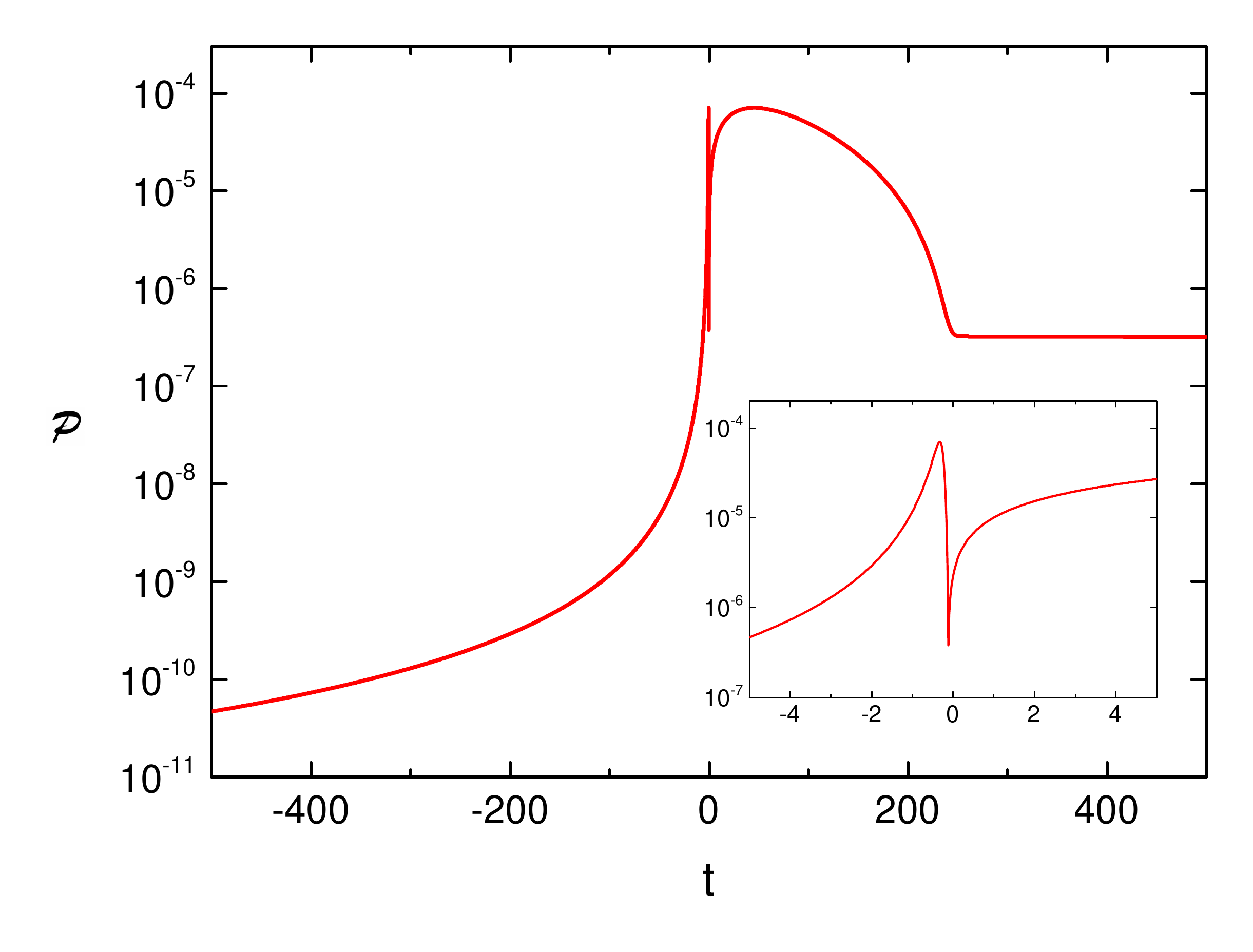} \label{fig:bgpicompare}}
\caption[]{\footnotesize \hangindent=10pt
\it{Evolution of the DHOST field $\pi$(represented by the solid orange curve), and its time derivative $\dot{\pi}$(represented by the solid purple curve), as a function of cosmic time $t$. We also plot out the function $\mathcal{P}$, whose positive behavior is required to stabilize the dynamics of the DHOST field $\pi$, as discussed in the last paragraph of Subsection \ref{subsec:bgeom}.}
}
\label{fig:bgfield}
\end{figure*}

It is hard to analytically solve Eq. \eqref{eq:Friedmann1} and Eq. \eqref{eq:Friedmann2}, so we studied the background dynamics numerically. We choose a specific value of model parameters as 
$\Lambda = 1 , \beta = 3.9, f = 10^{-2} $ (thus $\gamma = 10^{-6}$), $d_1=0.004$ and $d_2=0.18$
. The initial condition for the Galilean field is imposed to be the quasi-Minkowski solution described by Eq. \eqref{eq:pi} in the limit $t\rightarrow -\infty$. Our numerical results are presented in Fig. \ref{fig:bggeometry} and Fig. \ref{fig:bgfield}. All dimensional parameters are plotted in units of the reduced Planck mass $M_p \equiv 1$, and the horizontal axis denotes the cosmic time $t$. The numerical results of the background geometry in Fig. \ref{fig:bggeometry} show a pre-emergent phase before $t=0$. The EoS parameter of the DHOST field $w$ becomes $1/3$ at the late time, which implies that the emergent universe gracefully exits to a radiation dominated universe. The positivity of $\mathcal{P}$, which labels the stability of the background evolution of the DHOST field, is also illustrated in Fig. \ref{fig:bgfield}.

From Fig. \ref{fig:bggeometry}, we see that the background dynamics can be separated into three periods.
Before the emergent event $t=0$, $w$ approaches to minus infinity, and the scale factor $a$ is almost constant.
After the emergent event, $w$ experience a rapid change and approximately becomes constant for a short time.
Finally, after some certain time $t_B$, the scalar field becomes much larger than $1$, and the DHOST action is negligible compared to the Galileon part. This happens because $\dot{\pi}$ is comparably small, so the matter part of Eq. \eqref{eq:bounceS} returns to an ordinary matter, and $w$ is approximately one-third.

Based on the above observations, we make the following parametrization on the dynamics of the background geometry:
\[ w = \left\{ 
   \begin{array}{rl} 
   -\infty & \text{if } t < 0,\\ 
   1 & \text{if } 0 < t < t_B,\\ 
   1/3 & \text{if } t > t_B. 
   \end{array}  
 \right. \]
For a perfect fluid with constant $w$, the scale factor evolves as
\begin{equation}
\label{eq:atparametrization}
	a(t) \simeq t^p ~,~ p \equiv \frac{2}{3(1+w)} ~.
\end{equation}

There should be no infinity for EoS parameter $w$ in real physics, so we need to explain $w \simeq -\infty$ more carefully. If we substitute the approximate solution from Eq. \eqref{eq:pi} into Eq. \eqref{eq:Friedmann1} and Eq. \eqref{eq:Friedmann2}, we find $p_{\pi} = -(2+\beta)\gamma/t^4 + \mathcal{O}(t^{-6})$ and $\rho_{\pi} = \frac{(2+\beta)^2 \gamma^2}{12 t^6} + \mathcal{O}(t^{-8})$. This allows us to give an asymptotic behavior of $w$  and $\epsilon_H$ at $t \to -\infty$
\begin{equation}
\label{eq:winfty}
	 w=-\frac{12}{\gamma (2+\beta)} t^2 ~,~\epsilon_H= -\frac{18}{\gamma (2+\beta)} t^2 ~.
\end{equation}

The asymptotic behavior of the background parameter at $t \to -\infty$, to leading order is
\begin{equation}
\label{eq:bgparameterpast}
	\dot{H} = \frac{(2+\beta) \gamma}{2t^4} ~,~ H = - \frac{(2+\beta) \gamma}{6t^3} ~,~ \mathcal{P} = \frac{3\beta \gamma}{t^2} ~.
\end{equation}

\section{Perturbation analysis} \label{sec:Fluctuations}

This section is organized as follows. In Subsection \ref{subsec:perturbationgeneral}, we analyze the general property of scalar perturbation and confirm that both ghost and gradient instabilities are absent in our model. In Subsection \ref{subsec:perturbationdynamics}, we deduce the dynamical equation for the scalar perturbation and discuss its general properties. We consider two different mechanisms for the primordial cosmological perturbations, the scalar perturbations generated by vacuum fluctuations in Subsection \ref{subsec:vacuumdynamics} and by thermal fluctuations in Subsection \ref{subsec:thermal}, respectively, and summarize the corresponding scalar power spectrum in \ref{subsec:spectrum}.

For completeness, we also study the tensor perturbations in our model in appendix \ref{app:tensorpt}. We show that the tensor perturbations are stable (i.e., no ghost mode and no gradient instability), but may encounter the superluminality problem.

\subsection{General analysis on scalar modes and the stability check} \label{subsec:perturbationgeneral}

We study the perturbation theory of the present model by applying the ADM decomposition 
\begin{align*}
 ds^2 = N^2dt^2 - h_{ij}(dx^i+N^idt)(dx^j+N^jdt)~,
\end{align*}
where $N$ and $N^i$ are the lapse function and shift vector, respectively. 

We restrict our interest in scalar perturbations, and analysis on tensor perturbations is presented in appendix \ref{app:tensorpt}. For simplicity, we choose the uniform field gauge with $\delta\pi=0$ and $h_{ij} = a^2 e^{2\zeta}\delta_{ij}$. Now $\zeta$ represents the propagating scalar degree of freedom. We have already acquired the form of quadratic action for linear perturbations in our previous work \cite{Ilyas:2020qja} as
\begin{align}
\label{eq:S2scalar}
S_{2,s} 
= \int d \tau d^{3} x \frac{z_s^{2}}{2}  \Big[ \zeta'^2-c_{s}^{2}\left(\partial_{i} \zeta \right)^{2} \Big] ~,
\end{align}
where $z_s^2 $ and $ c_s^2$ take the form
\begin{align}
\label{eq:zs2}
z_s^2 = \frac{a^2\dot\pi^2{\cal P}}{(\gamma \dot\pi^3 - H)^2} ~,
\end{align}
and 
\begin{align}\label{eq:cs2}
(-\frac{z_s^2}{2a^2})c_s^2 = 1 + h + \frac{1}{a} \frac{d}{dt}  \bigg[ \frac{a \big( h_{X} \dot{\pi}^{2} - h - 1 \big) }{ H - \gamma \dot{\pi}^{3} } \bigg] ~,
\end{align}
where we use the fact $h_{\pi} \equiv (\partial h)/(\partial \pi) = 0$.

\begin{figure*}[th]
\centering
\subfloat[]
{\includegraphics[width=0.45\textwidth]{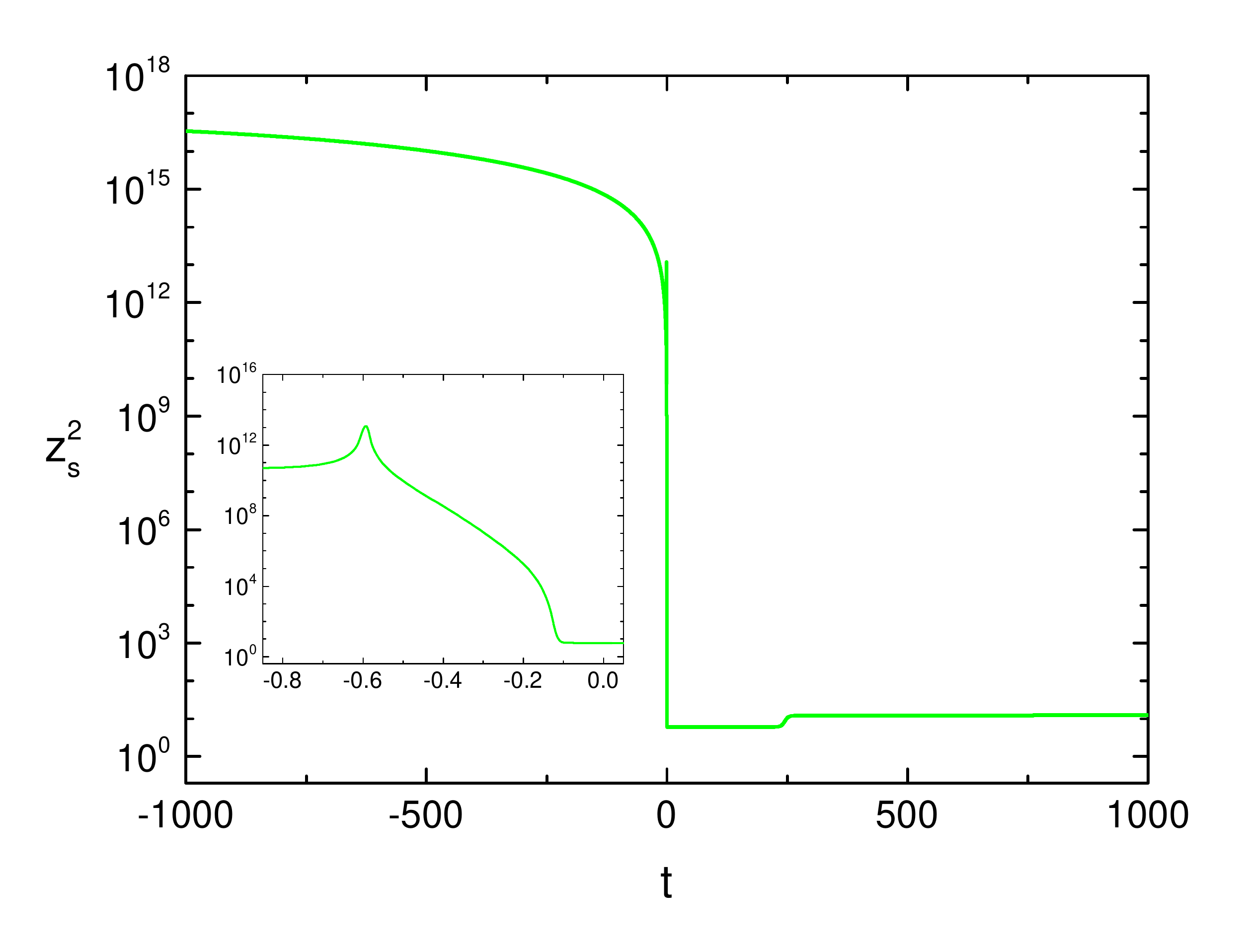} \label{fig:ptzs2}}
\subfloat[]
{\includegraphics[width=0.45\textwidth]{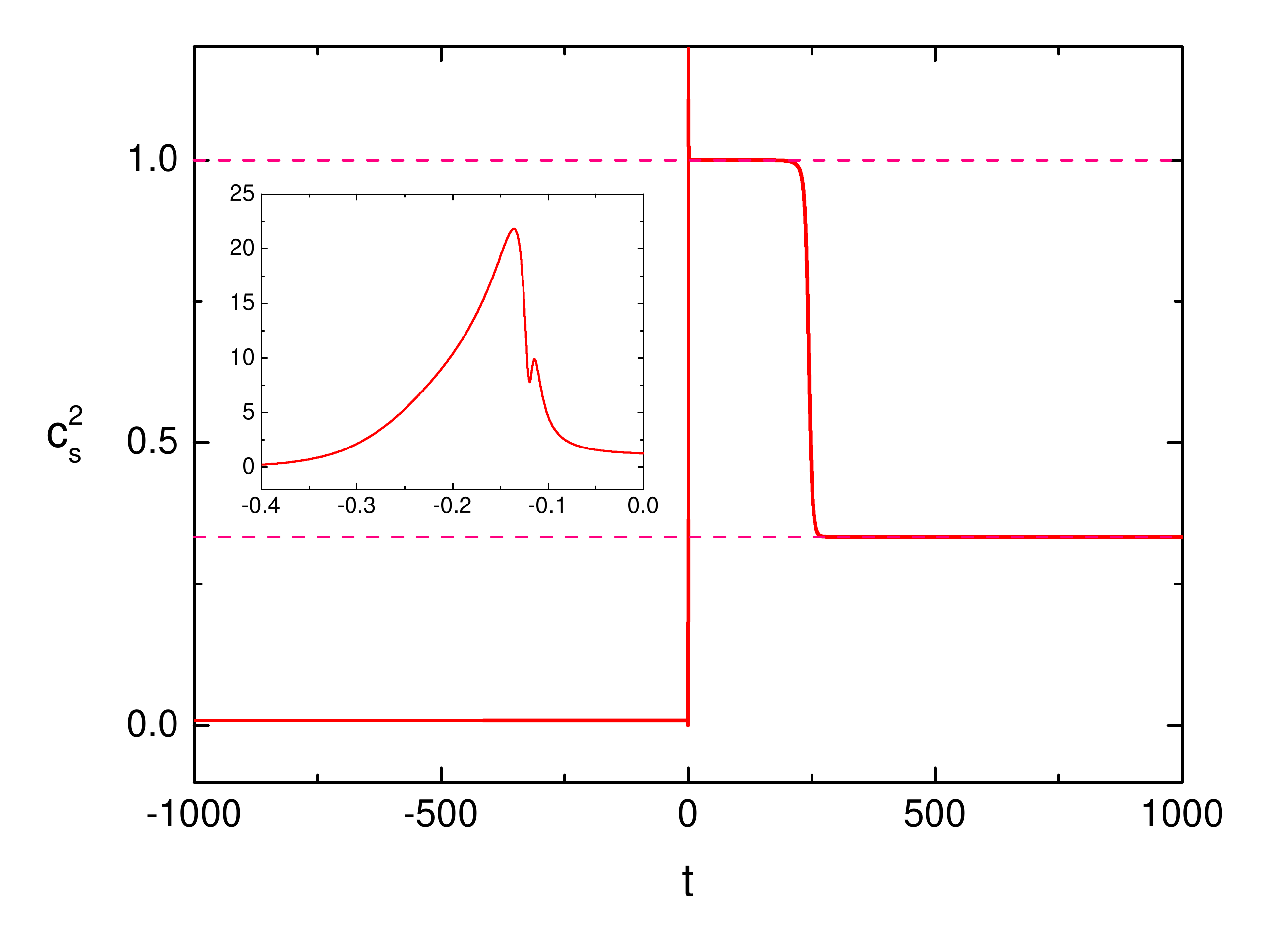} \label{fig:ptcs2}}
\caption[]{\footnotesize \hangindent=10pt
\it{The dynamics of $z_s^2$(the solid green curve in the left panel) and $c_s^2$(the solid red curve in the right panel) as a function of cosmic time $t$. The positivity of $z_s^2$ and $c_s^2$ is explicitly illustrated. We also show the approximated behavior of $c_s^2$, which equals $1$ shortly after the genesis and turns to $1/3$ at the far future by a pink dash line.
}
}
\label{fig:pt}
\end{figure*}

It is easy to see that the positivity of $\mathcal{P}$, which has been analyzed in the previous section, can lead to the positivity of $z_s^2$, so the ghost problem $z_s^2 < 0$ is absent. The form $c_s^2$ is more complicated than $z_s^2$, so we directly examine its positivity by numerical analysis in Fig. \ref{fig:pt}. From the numerical result, we see that $z_s^2$ and $c_s^2$ are strictly positive. Hence the gradient instability and ghost instability are absent in our model. The remaining issue is the superluminal propagation of DHOST field $\pi$ as $c_s^2$ would exceed unity in the vicinity of $t=0$, as shown in Fig. \ref{fig:ptcs2}. We shall discuss this issue in the conclusion section and appendix \ref{app:tensorpt}.

\subsection{Dynamics of scalar perturbations} \label{subsec:perturbationdynamics}

We follow the standard way to evaluate the power spectrum of late-time cosmological perturbations, as previously applied to inflationary cosmology \cite{Bardeen:1983qw, Brandenberger:1983tg}, string gas cosmology \cite{Nayeri:2005ck, Brandenberger:2006vv}, and bounce cosmology \cite{Cai:2009rd}. For linear scalar perturbations, we can get the dynamical equation by varying Eq. \eqref{eq:S2scalar} with respect to scalar perturbation parameter $\zeta$
\begin{align}
\label{eq:MSeqcosmic}
 \zeta_k^{\prime \prime} + \frac{2 z_s^{\prime}}{z_s} \zeta_k^{\prime} + c_s^2 k^2 \zeta_k = 0 ~,
\end{align}
where $\zeta_k$ is the Fourier-transformed $\zeta$, representing the scalar perturbation mode with wavenumber $k$. Defining Mukhanov-Sasaki (MS) variable $v_k = z \zeta_k$, we get the standard MS equation \cite{Sasaki:1983kd,Kodama:1985bj, Mukhanov:1988jd},
\begin{equation}
\label{eq:MSeq}
	v_k^{\prime \prime} + \left( c_s^2k^2 - \frac{z_s^{\prime \prime}}{z_s} \right) v_k = 0 ~.
\end{equation}

In the standard treatment, we separate the solution of Eq. \eqref{eq:MSeq} into two different scales. One is the sub-Hubble scale when the matter fluctuation part $c_s^2k^2$ dominates over the metric part $z''/z$, and second one is the super Hubble scale in which the metric dependent term $z''/z$ becomes dominant over term $c_s^2k^2$. 

We firstly consider the horizon crossing. Conventionally, the condition for Hubble crossing of a wave mode $k$ is
\begin{equation}
\label{eq:Hubblecrossing}
	kc_s \simeq a(t_H)H(t_H) ~,
\end{equation}
where $t_H \equiv t_H(k)$ denotes the cosmic time of Hubble-crossing for a given wavenumber $k$. We recover $c_s$ in Eq. \eqref{eq:Hubblecrossing} since $c_s^2$ is not always equal to unity in our model. 

\begin{figure*}[th]
\centering
\subfloat[]
{\includegraphics[width=0.45\textwidth]{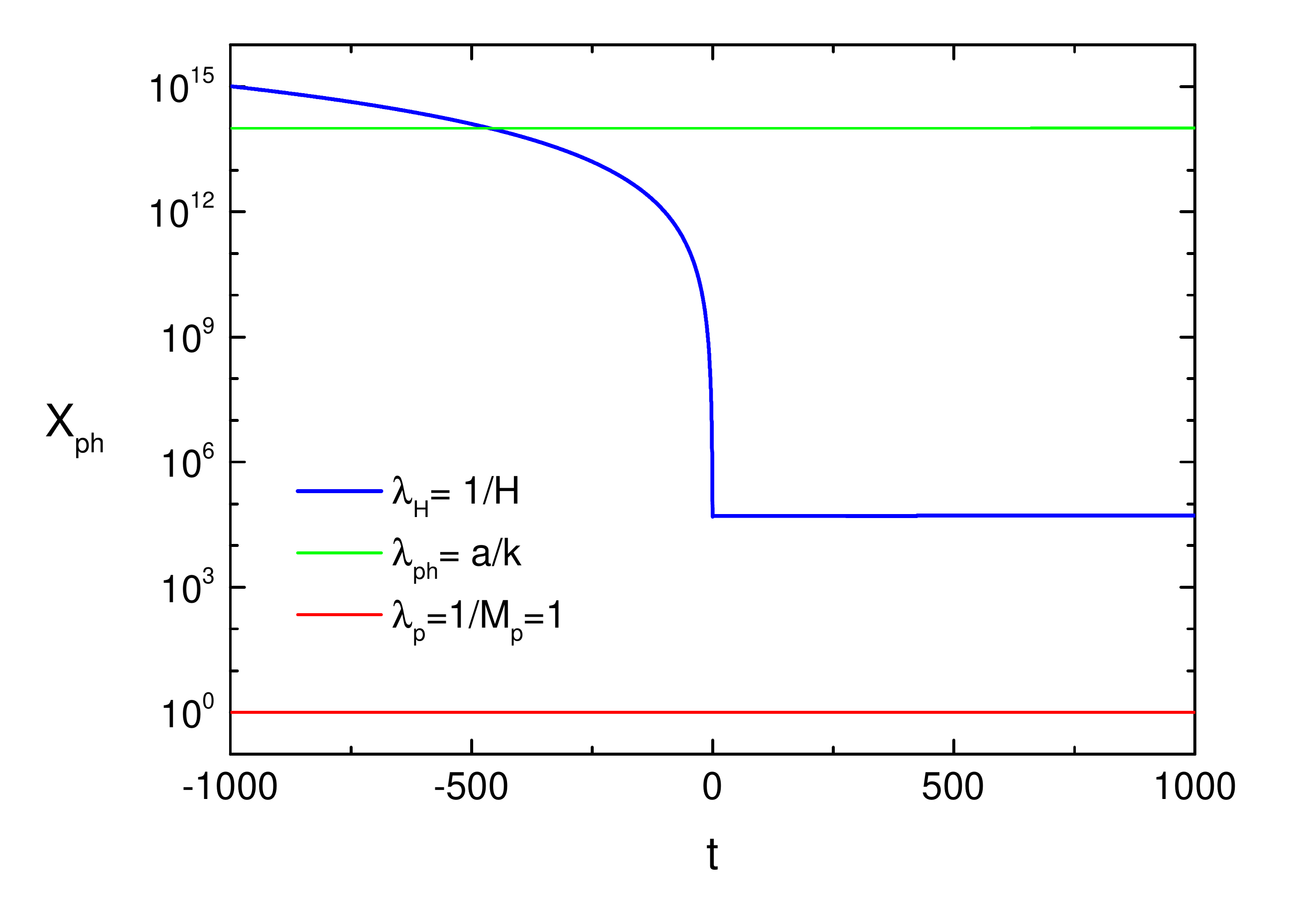} }
\caption[]{\footnotesize \hangindent=10pt
\it{A sketch of the evolution of length scales in the emergent universe scenario. The vertical axis is the physical spatial coordinate $X_{ph}$, and the horizontal axis is the cosmic time $t$. The physical wavelength $\lambda_{ph} = a/k$ of the mode with comoving wavenumber $k$ is depicted in solid green line, while the Hubble radius $\lambda_H = |H|^{-1}$ is depicted in the blue curve. We also draw the Planck length $\lambda_p = M_{pl}^{-1}$ in the red line for comparison. The wavenumber is taken to be $k = 10^{-14}$.
}
}
\label{fig:horizon}
\end{figure*}

We plot the evolution of the length scales in figure \ref{fig:horizon}, in which we take a special wavenumber $k=10^{-14}$ and $c_s = 1$. The numerical results show that the horizon crossing happens in the pre-emergent phase. This is generic for both vacuum fluctuation and thermal fluctuation in our model: the Hubble radius becomes almost constant shortly after the emergent event with a small value $\lambda_H = \mathcal{O} (10^{5})$, as we see from Fig. \ref{fig:horizon}. Compared to the upper limit of the observed wavenumber $k_{\max} = \mathcal{O}(10^{-8})$, which indicates a physical wavelength larger than $a/k_{\max} = 10^8$, we see that all perturbation modes become super-horizon at the emergent event. Hence in our model, the horizon crossing should happen in the pre-emergent phase $t<0$.

Secondly, we consider the super-Hubble case, where the dynamics is determined mainly by the background geometry. It is well-known that, for a cosmological background filled with a matter of constant EoS parameter $w$, the general solution to Eq. \eqref{eq:MSeq} on the super-Hubble scale is
\begin{align*}
	\zeta_k = D + \frac{S}{(\eta)^{\nu}} ~,
\end{align*}
where $D$ and $S$ are constants determined by the matching condition, and $\nu \equiv 3(1-w)/(1+3w)$. It is clear that after horizon crossing, there propagate two linearly independent modes. The constant mode is denoted as ``$D$-mode'', and the decreasing mode is denoted as ``$S$-mode''. Conventionally, it is assumed that the D-mode dominates the scalar power spectrum since the S-mode will vanish as the universe expands. So we shall consider the scale invariance generated by the D-mode only.

Finally, we discuss the matching conditions for cosmological perturbations, where the equation of state $w$ undergoes a sharp jump. We follow the technique developed in \cite{Hwang:1991an, Deruelle:1995kd}, which indicates that $v_k$ and $v_k^{\prime}$ are continuous across the matching surfaces, i.e. $t=0$ and $t = t_B$. In other words, we can ignore the effect of the matching process on the scalar perturbation $v_k$, and hence the scalar power spectrum.

\subsection{Vacuum fluctuations} \label{subsec:vacuumdynamics}

In this subsection, we consider the case where the primordial perturbation comes from the quantum vacuum fluctuations of the DHOST field $\pi$ on sub-Hubble scales. In principle, we should study the dynamical equation of $v_k$ in each period, then connect the perturbation by appropriate matching conditions and finally study the scalar power spectrum when the perturbation becomes super-Hubble scale. However, a simple observation from Fig. \ref{fig:bggeometry} reveals that $a(t) \simeq 1$ and $H \simeq 2 \times 10^{-5}$ after the emergent event. Since the typical co-moving wavelength of observed scalar perturbation is $k \ll 10^{-8}$, we see $aH \gg k$ when $t>0$, and the perturbation of observational interest is super-Hubble scale. According to the above analysis, it is sufficient to take the scalar perturbation $\zeta_k$ as constant after the horizon crossing. Then we only need to consider the dynamics of scalar perturbation in pre-emergent phase.

The action for the linear perturbation in Eq. \eqref{eq:S2scalar} is that of a harmonic oscillator, so we impose the quantum vacuum initial conditions
\begin{equation}
\label{eq:vkinitial}
    v_k(\eta) \to \frac{e^{-ic_s k(\eta - \eta_0)}}{\sqrt{2k}} ~,
\end{equation}
where $\eta_0$ is some constant representing the freedom of choosing initial phase.

At the far past, we can get the function $c_{s,0}^2$ and $z_s^2$ with the help of the asymptotic behavior from Eq. \eqref{eq:bgparameterpast} and Eq. \eqref{eq:pi}, and use the fact that the conformal time approximately equals to the cosmic time $t$ since $a$ is approximately equal to $1$, which reads
\begin{align}
\label{eq:zs2farpast}
	c_{s,0}^2\sim \frac{4-\beta}{\beta}\simeq 0.0085,\\~z_s^2 \sim \frac{108\beta }{(\beta - 4)^2 \gamma} a^2\eta^2 ~, z_s \propto a\eta ~.
\end{align}
We find that $c_s^2$ remains almost constant in the pre-emergent phase, which can also be read from Fig. \ref{fig:pt}.

As the time approaches $t\sim -2$, the approximation breaks down. However, the scalar perturbation will have an observational window corresponding to 15 effective e-flods \cite{Akrami:2018odb}, which is much larger than the time interval $-2<t<0$. So the shape of the observed scalar power spectrum cannot be influenced in this short period. Hence in this paper, we will ignore the break down of the approximation \eqref{eq:zs2farpast} during the short time interval.

The background effect appearing in the MS equation is
\begin{align}
\label{eq:zs2parat<0}
    \frac{z_s^{\prime \prime}}{z_s} = \frac{2H}{\eta} + H^2 + H^{\prime} \simeq \frac{(2+\beta)\gamma}{6\eta^4} ~,
\end{align} 
Then, Eq. \eqref{eq:MSeq} takes the form as
\begin{equation}
\label{eq:MSeqeta4}
	v_k^{\prime \prime} + \left( c_{s,0}^2 k^2 - \frac{(2+\beta)\gamma}{6\eta^4} \right) v_k = 0 ~.
\end{equation}
At far past, the dominant term in the parentheses of Eq. \eqref{eq:MSeqeta4} is $c_{s,0}^2k^2$. For a certain conformal wavenumber $k$, the term $(2+\beta)\gamma / (6\eta^4)$ becomes comparable to $c_{s,0}^2k^2$ at the time 
\begin{equation}
    \eta_k = - \left[ \frac{(2+\beta) \gamma}{6 c_{s,0}^2k^2} \right]^{\frac{1}{4}}  \to  \eta_k \leq -1000 ~,
\end{equation}
where the upper bound is taken when $k$ takes the upper limit of observed wavelength $k_{\max} = 10^{-8}$. This again verifies the argument in Subsection \ref{subsec:perturbationdynamics} that the perturbation should cross the horizon at $t<0$.

The dynamics of the perturbation can be approximately separated into two parts. For each wave mode $k$, the perturbation $v_k$ will evolve with the dominance of $c_{s,0}^2k^2$ until $\eta = \eta_k$, then it crosses the horizon and becomes almost constant. The general solution to Eq. \eqref{eq:MSeqeta4} is 
\begin{equation}
\label{eq:vkt<0}
	v_k(\eta) = C_{0,+} e^{i c_{s,0}k \eta} + C_{0,-} e^{-i c_{s,0}k \eta} ~,
\end{equation}
where $C_{0,+}$ and $C_{0,-}$ are integration constants. One may comprehensively understand the Eq. \eqref{eq:vkt<0} by noticing that, in the pre-emergent phase, the background geometry is quasi-Minkovskian, so the background should contribute little to the perturbation dynamics, and $v_k$ propagates as a plane wave.

Imposing the initial condition from Eq. \eqref{eq:vkinitial} to the general solution in Eq. \eqref{eq:vkt<0}, $C_{0,+}$ and $C_{0,-}$ should have $k^{-\frac{1}{2}}$ dependence. At the Hubble crossing for a fixed $k$, the k-dependence of the scalar perturbation $v_k$ will be
\begin{equation}
 v_k(\eta_k) \propto k^{-\frac{1}{2}} e^{2ic_{s,0}k(\eta_k-\eta_{k,c})} ~,
\end{equation}
where $\eta_{k,c}$ represents a phase factor. The scalar power spectrum becomes
\begin{equation}
\label{eq:ps}
 P_{\zeta} \propto k^3 \bigg| \frac{v_k(\eta_k)}{z(\eta_k)} \bigg| ^2 \propto k^{3} ~,
\end{equation}

The expression for the scalar power spectrum in Eq. \eqref{eq:ps} tells that the vacuum fluctuation of the DHOST field itself cannot generate a scale-invariant power spectrum. This is a common feature in the original Galileon Genesis models, who fails to produce scale-invariant curvature perturbations without invoking the curvaton mechanism \cite{Creminelli:2010ba, Creminelli:2012my, Hinterbichler:2012fr, Hinterbichler:2012yn} unless one generalize the Galileon Genesis model at the cost of producing a strongly blue spectra of tensor perturbation \cite{Cai:2014uka, Nishi:2015pta, Nishi:2016wty}. This feature also happens in other emergent universe scenarios. For example, in the emergent universe scenario constructed by quintom matter, it is found that the matter field will remain in the vacuum state and hence the scalar power spectrum is always blue \cite{Cai:2012yf}, unless introducing a curvaton with specific kinetic coupling to the original matter field \cite{Cai:2013rna}. It is then interesting to investigate whether one can impose curvaton mechanism or improve the model to find a mechanism of generating a scale-invariant primordial power spectrum of curvature perturbation in follow-up work.

\subsection{Thermal fluctuations} \label{subsec:thermal}

In this subsection, we consider the case where the primordial perturbation is generated by fluctuations from a thermal gas. We assume that the thermal gas has negligible contribution to the background dynamics, then we only need to consider its influence on perturbation. 

We begin with the case where the thermal gas is composed of point particles with an arbitrary EOS $w_r$. The thermal fluctuation has been studied previously in string gas cosmology \cite{Brandenberger:2006vv} and bounce cosmology \cite{Cai:2009rd}. In each case, the scale invariance can be possibly generated by the thermal fluctuation mechanism. We will closely follow the methods in the above two papers.

From the stress-energy conservation equation, it follows that the energy density and temperature of the thermal gas change with a function of the scale factor as
\begin{equation}
\label{eq:rhoT}
 \rho_r \simeq a^{-3(1+w_r)} ~, T \simeq a^{-3w_r} ~, 
\end{equation}
Now we can get the heat capacity of the gas by
\begin{equation}
\label{eq:Cv}
 C_V(R) = R^3 \frac{\partial \rho}{\partial T} \simeq R^3 T^{\frac{1}{w_r}} ~,
\end{equation}
where $R \equiv R(k)$ is the physical length corresponding to the co-moving momentum scale $k$, and we may take it as the Hubble radius $R \sim 1/H$.
Up to a constant of order $\mathcal{O}(1)$, the power spectrum of the metric perturbation at horizon crossing moment, generated by the thermal gas \cite{Cai:2009rd}, is
\begin{equation}
\label{eq:Pthermalgeneral}
 P_{\Phi}(k)(t_H(k)) = \frac{1}{4} \frac{C_V(R)}{H^4_{t_H(k)}} \frac{T^2}{R^6} ~,
\end{equation}
where $H_{t_H(k)}$ is the Hubble parameter at horizon crossing. 
Combining Eq. \eqref{eq:rhoT}, Eq. \eqref{eq:Cv} with Eq. \eqref{eq:Pthermalgeneral}, and applying the Hubble crossing condition $k=aH\simeq H$, we get 
\begin{equation}
 P_{\Phi}(k)(t_H(k)) \sim k^{-1} ~.
\end{equation}
Making use of the relation between gravitational potential $\Phi$ and $\zeta_k$
\begin{equation}
    \Phi_k=-\frac{a^2 \dot{H}}{k^2H}\dot{\zeta},
\end{equation}
and the fact that $v_k$ oscillates with frequency $k/a$ on sub-Hubble scales, one has
\begin{equation}
\label{phitozeta}
    P_\zeta(k)(t_H(k))=\frac{k^2}{a^2H^2\epsilon_H^{~2}}P_\Phi\sim k^{\frac{4}{3}}P_\Phi.
\end{equation}
In the last step we have used the background equation from Eq. \eqref{eq:winfty}. Thus, we obtain
\begin{equation}
    P_\zeta(k)(t_H(k))\sim k^{\frac{1}{3}} ~,
\end{equation}
which gives a blue tilted spectrum.

We then discuss two more nontrivial cases where the thermal dynamics is governed by string theory. The first case is the Gibbons-Hawking (GH) radiation \cite{Gibbons:1977mu}, which has been studied extensively in the context of developments in string theory, particularly in light of the role of holography \cite{Fischler:1998st}. the heat capacity for a thermal GH radiation is
\begin{equation}
\label{eq:CvGH}
	C_V(R) \sim R^2,
\end{equation}
which is derived from the average energy of GH radiation $\left\langle E \right\rangle = TR^2$. Combining Eq. \eqref{eq:Pthermalgeneral} with Eq. \eqref{eq:CvGH} and use the definition of the Gibbons-Hawking temperature associated with the instantaneous Hubble radius $T = 1/R \sim H$, the scalar power spectrum is
\begin{align}
	P_{\Phi}(k) \sim k^{2}.
\end{align}

With the help of Eq. \eqref{phitozeta}, we obtain 
\begin{align}
	P_{\zeta}(k) \sim k^{\frac{10}{3}}.
\end{align}

The second case corresponds to the thermal fluctuation from the string gas. As revealed in string gas cosmology \cite{Nayeri:2005ck, Brandenberger:2006vv, Brandenberger:2008nx, Brandenberger:2011et}, the gas of closed strings induces a scale-invariant spectrum with a slight red tilt of scalar metric fluctuations on all scales smaller than the Hubble radius, as long as the fluctuations exit the Hubble radius at the end of a quasi-static Hagedorn phase.

\subsection{Scale invariance of power spectrum} \label{subsec:spectrum}

We have considered primordial perturbations originated from the vacuum and thermal fluctuations. The results are summarized as follows:
\begin{equation}
 P_{\zeta}=\left\{\begin{array}{ll}k^{3}, & \text {vacuum fluctuations} \\ 
 k^{1/3}, & \text { thermal particle fluctuations} \\ k^{10/3}, & \text{Gibbons-Hawking  radiation} \\ k^{0_-}, & \text{string gas}\end{array} \right.
\end{equation}
where the symbol $k^{0_-}$ represents a scale-invariant spectrum with a slight red tilt. The results show that thermal fluctuations from a string gas can provide a scale-invariant power spectrum, which is a generic feature of string gas cosmology \cite{Nayeri:2005ck, Brandenberger:2006vv, Brandenberger:2008nx, Brandenberger:2011et}. The other situations always give a blue scalar power spectrum. We will extend our model in the future work to generate a nearly scale-invariant spectrum with mechanisms other than the string gas paradigm.

\section{Conclusion and discussions} \label{sec:conclude}

In this paper, we present a realization of the emergent universe scenario by introducing a deformed kinetic term and a DHOST coupling to the original Galileon Genesis model. We first investigate the cosmological evolution and show that the universe can gracefully exit the emergent phase and transfer to a radiation dominated phase. The model can meet the standard thermal history of the universe without additional mechanisms like the decay of the Galileon field into radiation. We then study the cosmological perturbation and show that the model is free from gradient instability problem. 

We have also investigated various theoretical conditions for deriving a nearly scale-invariant power spectrum for primordial scalar perturbations. We find that the scale invariance cannot be generated in our model through the vacuum fluctuation of the DHOST field. Fluctuations from thermal string gas can generate a scale-invariant power spectrum with a slightly red tilt, while that from thermal gas of point particles and GH radiation will always result in a blue spectrum. It would be interesting to see whether a scale-invariant primordial power spectrum of curvature perturbation can be achieved with curvaton mechanism \cite{Rubakov:2009np, Wang:2012bq} or improving the current model. It is also crucial to confront our model with various cosmological constraints, including the high precision CMB measurement of the primordial power spectrum, the tensor-to-scalar ratio, primordial non-gaussianities, and so on. All these topics are to be studied in the following-up work.

Moreover, the sound speed square of both scalar and tensor perturbations will be larger than unity near the emergent event, triggering the superluminality problem. It has been pointed out that this issue is a generic property for cosmological applications of Galilean/Horndeski theories \cite{Mironov:2020mfo, Bruneton:2006gf, Babichev:2007dw}. We argued that for the DHOST cosmology, this issue is quite model-dependent \cite{Ilyas:2020qja}. Thus, we expect careful model construction of DHOST Genesis may avoid the superluminality issue.

We end by commenting that it is essential to study the particle production process for the universe depicted by our model, which will be addressed in a follow-up study.

\section*{Acknowledgments}
We are grateful to Robert Brandenberger, Damien Easson, Xian Gao, Emmanuel Saridakis, Misao Sasaki, Dong-Gang Wang and Yi Wang for stimulating discussions. 
This work is supported in part by the NSFC (Nos. 11722327, 11961131007, 11653002, 11847239, 11421303), by the CAST Young Elite Scientists Sponsorship (2016QNRC001), by the National Youth Talents Program of China, by the Fundamental Research Funds for Central Universities, by the USTC Fellowship for international students under the ANSO/CAS-TWAS scholarship, and by GRF Grant 16304418 from the Research Grants Council of Hong Kong. 
Amara Ilyas and Mian Zhu contributed equally to this work.
All numeric are operated on the computer clusters {\it LINDA \& JUDY} in the particle cosmology group at USTC.

\appendix

\section{A brief introduction to DHOST theory}
\label{app:DHOSTintro}

In this appendix, we give a brief introduction to the DHOST theory. DHOST theories are defined to be the maximal set of scalar-tensor theories in four dimensional space-time that contain at most three powers of second derivatives of the scalar field $\pi$, while propagating at most three degrees of freedom , and Galileon theory is the specific case of DHOST theory where EoM of the scalar field remains second order. 

The most general DHOST action involving up to cubic powers of second derivative of the scalar field $\pi$ can be written as 
\begin{eqnarray}
\label{dhost32}
S[g,\pi]& = \int d^4x \sqrt{-g} \Big[ h_2(\pi, X) R +  C^{\lambda \nu \rho \delta}_{(2)} \pi_{\lambda \nu} \pi_{\rho \delta}  \nonumber \\ 
& + h_3(\pi, X) G_{\lambda \nu} \pi^{\lambda \nu} +C^{\lambda \nu \rho \delta \alpha \beta}_{(3)} \pi_{\lambda \nu} \pi_{\rho \delta} \pi_{\alpha \beta} \Big] ~.
\end{eqnarray} 
The tensors $C_{(2)}$ and $C_{(3)}$ represent the most general tensors constructed with the metric $g_{\lambda \nu}$ as well as the first derivative of the scalar field,  which is denoted as $\pi_{\lambda} \equiv \nabla_{\lambda} \pi$. The symbol $\pi_{\lambda \nu}$ denotes the second derivative $\pi_{\lambda \nu} \equiv \nabla_{\lambda} \nabla_{\nu} \pi$,and the canonical kinetic term $X$ is defined as $X \equiv \frac{1}{2} \nabla^{\nu} \pi \nabla_{\nu} \pi$. Exploiting the symmetry in Eq. $C_{(2)}$ and Eq. $C_{(3)}$, one can reformulate Eq. \eqref{dhost32} to be
\begin{equation}\nonumber
C^{\lambda \nu \rho \delta}_{(2)} \pi_{\lambda \nu} \pi_{\rho \delta} +  C^{\lambda \nu \rho \delta \alpha \beta}_{(3)} \pi_{\lambda \nu} \pi_{\rho \delta} \pi_{\alpha \beta} \equiv \sum_{i=1}^5 a_i L_i^{(2)} + \sum_{j=1}^{10} b_j L_j^{(3)} ~,
\end{equation}
where $a_i$'s and $b_i$'s depend on $\pi$ and $X$. We only need the quadratic DHOST terms, so we consider all $b_j$ terms to vanish in our model.

The Lagrangian coupling to $a_i$'s are defined as
\begin{align*} 
\label{eq:L_i^2}
& L_1^{(2)} = \pi_{\mu \nu} \pi^{\mu \nu} ~,~ L_2^{(2)} = (\Box \pi)^2 ~,~ L_3^{(2)} =  (\Box \pi)\pi^{\mu}\pi_{\mu \nu} \pi^{\nu} ~, \nonumber\\ 
& L_4^{(2)} = \pi_{\mu}\pi^{\mu \rho}\pi_{\rho \nu}\pi^{\nu} ~,~ L_5^{(2)} = (\pi^{\mu}\pi_{\mu \nu}\pi^{\nu})^2 ~,
\end{align*}
and to make the theory not propagating the ghost degree of freedom, the form of $a_i$'s are severely constrained. There are six possible combinations of $a_i$'s which can give a healthy action without ghost and for our purpose we shall concentrate on the type $^{(2)} N$\textendash$II$ DHOST theory, where there are three free functions $h_2$, $a_4$ and $a_5$, and the others are constrained by
\begin{equation}
\label{eq:dhostconstraint}
	a_2 = -a_1 = \frac{h_2}{2X} ~,~ a_3 = \frac{h_2 - 2 X h_{2X}}{2 X^2} ~.
\end{equation}

We can see from Eq. \eqref{eq:dhostconstraint} that the action in \eqref{eq:bounceS} is of the type $^{(2)} N$\textendash$II$ DHOST theory, so our model is free from Ostrogradsky instabilities.

\section{Tensor perturbations} \label{app:tensorpt}

In this appendix, we work out the tensor perturbations of our model. We will closely follow the technique developed in \cite{Gao:2019liu}. The quadratic action for tensor modes in the FLRW background takes the generic form as
\begin{equation}
\label{eq:S2tgeneral}
S_{2,T(General)} = \int dt d^3x \frac{a^3}{2} \left( \dot{\gamma}_{ij} \hat{\mathcal{G}}^{ij,kl}\dot{\gamma}_{kl} - \gamma_{ij} \hat{\mathcal{W}}^{ij,kl} \gamma_{kl} \right) ~,
\end{equation}
in which $\gamma_{ij}$ are tensor perturbations and $\hat{\mathcal{G}}$,$\hat{\mathcal{W}}$ are determined by the theory. Then, we can substitute Eq. \eqref{eq:bounceS} into Eq.\eqref{eq:S2tgeneral} and get 
\begin{equation}
\label{eq:S2tDHOST}
S_{2,T(DHOST)} = \int d\eta d^3x \frac{a^2}{8}\left[ \gamma_{ij}^{\prime 2} - (1+h)(\nabla_k \gamma_{ij})^2 \right] ~,
\end{equation}
where $\eta$ is the conformal time defined by $d\eta = dt/a$, and $\prime$ represents differentiation with respect to $\eta$. One can straightforwardly see from Eq. \eqref{eq:S2tDHOST} that the ghost problem is absent in our case. The propagation speed of tensor modes is expressed by $c_T^2 = 1+h$. In order to ensure the tensor modes of our model are free from gradient instability, the condition of $1+h > 0$ is required to be satisfied.

\begin{figure}[h]
\centering
\includegraphics[width=0.40\textwidth]{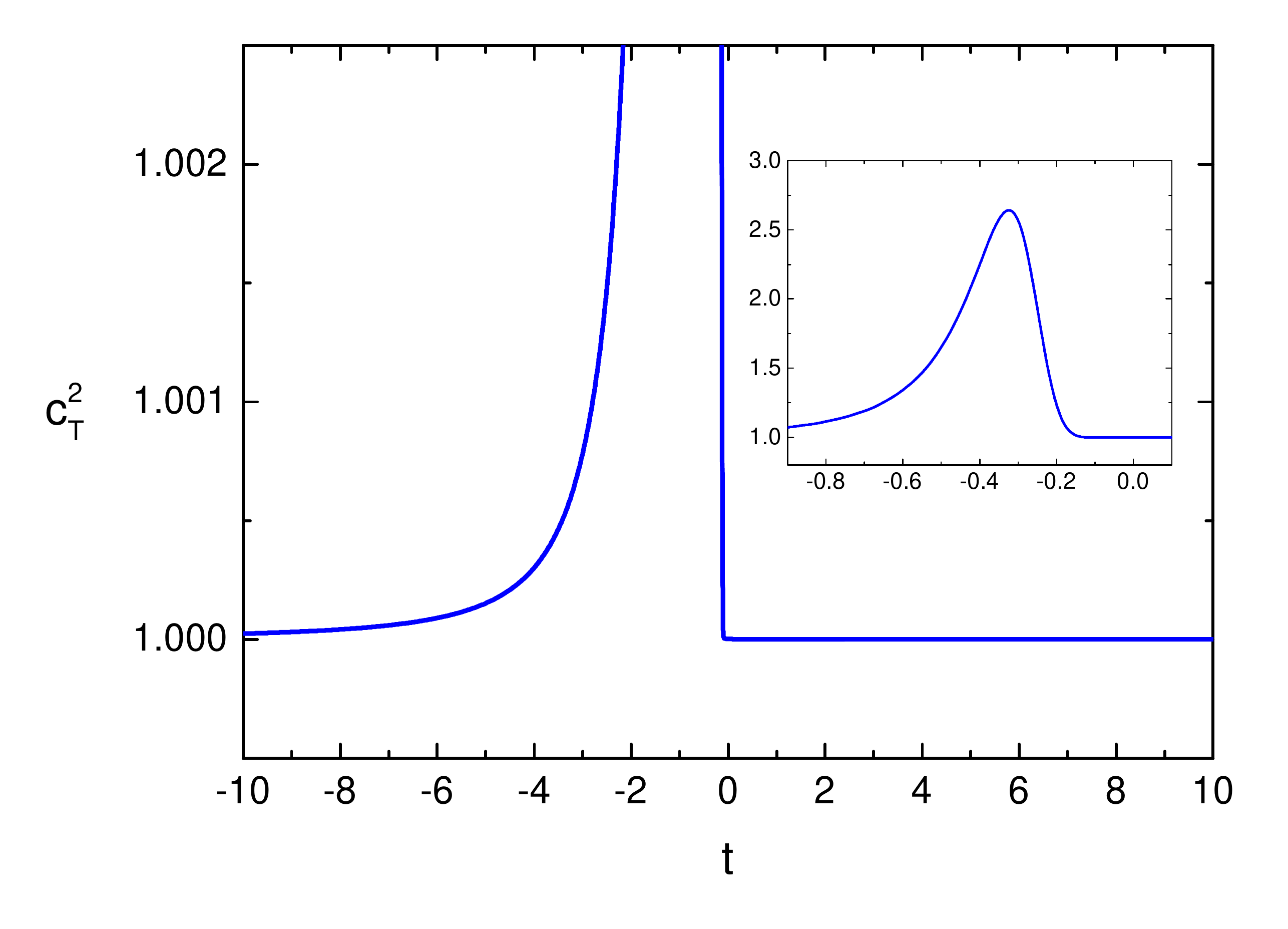}
\caption[]{\footnotesize \hangindent=10pt
\it{The dynamics of the sound speed of tensor perturbation $c_T^2$ as a function of cosmic time $t$. The positivity of $c_T^2$ is exhibited in the whole cosmological process, but $c_T^2$ will exceed $c^2 = 1$ in the neighborhood of $t=0$. }
}
\label{fig:ct2}
\end{figure}

We numerically plot the sound speed of tensor perturbation $c_T^2$ in Fig. \ref{fig:ct2}. We illustrate that $c_T^2$ is positive during the whole cosmological process, but it will exceed the speed of light squared, i.e., $c^2 > 1$ for a short time interval near the genesis event, triggering the superluminality problem. The existence of superluminality is a general feature in scalar-tensor theory beyond Horndeski \cite{Mironov:2020mfo}, and here we shall argue that superluminality does not necessarily correspond to acasuality \cite{Bruneton:2006gf, Babichev:2007dw, Kang:2007vs, Deffayet:2010qz, Dobre:2017pnt}. Still, it is interesting to see whether we can develop a genesis model without the superluminality issue.

\pagebreak

\end{document}